\documentclass[preprint]{aastex}

\tighten
\newcommand{\asec}{\hbox to 1pt{}\rlap{$^{\prime\prime}$}.\hbox to 2pt{}}
\newcommand{\amin}{\hbox to 1pt{}\rlap{$^{\prime}$}.\hbox to 2pt{}}
\newcommand{\sdeg}{\hbox to 1pt{}\rlap{$^{\circ}$}.\hbox to 2pt{}}

\shortauthors{Lauer \& Boroson}
\shorttitle{Observations of SDSS J153636.22+044127.0}

\begin{document}

\title{{\it HST} Images and KPNO Spectroscopy of the
Binary Black Hole Candidate SDSS J153636.22+044127.0\footnote{Based on observations made with the NASA/ESA
{\it Hubble Space Telescope}, obtained at the Space Telescope Science Institute,
which is operated by the Association of Universities for
Research in Astronomy, Inc., under NASA contract NAS 5-26555. These
observations are associated with GO/DD proposal \# 11993.}}

\author{Tod R. Lauer \& Todd A. Boroson}
\affil{National Optical Astronomy Observatory\footnote{The National Optical
Astronomy Observatory is operated by AURA, Inc., under cooperative agreement
with the National Science Foundation.},       
P.O. Box 26732, Tucson, AZ 85726}

\vfill

\begin{abstract}

We present {\it HST} WFPC2/PC images and KPNO 4-m longslit
spectroscopy of the QSO SDSS J153636.22+044127.0, which
we advanced as a candidate binary
supermassive black hole.  The images reveal a close companion
coincident with the radio source identified by \citet{wl}.
It appears to be consistent with a $M_g\sim-21.4$ elliptical galaxy,
if it is at the QSO redshift.
The spectroscopy, however, shows no spatial offset between
the red or blue Balmer lines.
The companion is thus not the source
of either the red or blue broad line systems;
SDSS J153636.22+044127.0 is highly unlikely to be a
chance superposition of objects or an ejected black hole.
Over the $\Delta T=0.75$ yr difference between
the restframe epoch of the present and SDSS spectroscopy,
we find no velocity shift to within 40 km s$^{-1}$,
nor any amplitude change in either broad line system.
The lack of a shift can be admitted under the binary hypothesis,
if the implied radial velocity is a larger
component of the full orbital velocity than was assumed in our earlier work.
A strong test of the binary hypothesis requires
yet longer temporal baselines.
The lack of amplitude variations
is unusual for the alternative explanation of this object
as a ``double-peaked'' emitter; we further argue that
SDSS J153636.22+044127.0 has unique spectral features that have
no obvious analogue with other members of this class.

\end{abstract}


\section{A Binary Supermassive Black Hole Candidate}

The low-redshift QSO SDSS J153636.22+044127.0 (hereafter J1536+0441)
is a candidate for hosting a binary supermassive black hole \citep{tnt}.
Its spectrum exhibits two broad-line emission systems
at $z=0.388$ and $z=0.373,$ thus
separated in velocity by 3500 km${\rm~s^{-1}}.$ but only one narrow line
system, which is associated with the redder or `r' broad-line system.
A third system of unresolved absorption lines has an intermediate velocity.
These characteristics are unique among known quasars.
We advanced this object as a candidate binary system of two
supermassive black holes.
The rough estimates of the two black hole masses
are $10^{8.9}M_\odot$ and $10^{7.3}M_\odot,$
based on line and continuum properties.
Under this picture, the velocity difference between the broad-line systems
is the projected orbital velocity difference of the two black holes.
Assuming a circular orbit with random phase and inclination,
we derive the separation of the black holes and their orbital period to be
$\sim0.1$ parsec and $\sim100$ years, respectively.

There is strong interest in the prevalence of binary black hole systems,
which should be commonly formed in the formation of galaxies from
hierarchical merging of smaller systems \citep{vhm}.
This has motivated several
observational searches for binary systems; however, few compelling
candidates have survived scrutiny.  Likewise, other investigators
have quickly attempted to obtain additional observations of
J1536+0441 to ascertain its viability as a binary system.  These
works, rather than confirming the binary hypothesis, have offered other
interpretations.

\citet{chorn} obtained a spectrum of J1536+0441 that
covered longer wavelengths than those that were sampled by the SDSS spectrum.
They identified a ``bump" in the red-wing of the r-system H$\alpha$
line that has a corresponding weak feature in the Mg II $\lambda$ 2800 line.
At H$\beta$, the presence of this red bump is ambiguous due
to its near overlap with the 4959\AA\ [O III] line.
Further, the red bump and b-system (bluer) Balmer lines are roughly symmetrical
about the r-system lines.
The red-bump, and in particular its apparent symmetry with the b-system,
is not anticipated in the binary black hole model.
\citet{chorn} instead suggest that J1536+0441 is a ``double-peaked"
emission-line QSO.
\citet{gas} also suggested this interpretation, based
on his evaluation of the SDSS spectrum, alone.
Under this picture, the spectrum of J1536+0441 reflects
instabilities in or an unusual configuration of an accretion disk
around a single black hole.  At the same time, \citet{chorn}
point out that the extreme difference in flux between the b-system
lines and the red-bumps, plus the sharp cores of the b-system Balmer lines,
are unique among the class of double-peaked emitters.

Other interpretations are that J1536+0441 is the chance superposition
of two AGN, or that one system is a black hole with associated AGN that
has been ejected from the nucleus of the host galaxy.  We showed,
however that there is only a $3\times10^{-3}$ chance of finding
any random close pair of QSOs in the entire SDSS DR7 sample
searched to yield J1536+0441.\footnote{This assumes an uncorrelated Poisson
angular distribution of QSOs, which is reasonable, since the broad-line
velocity difference, if interpreted as an offset in the Hubble flow,
corresponds to a larger physical separation in distance than any expected 
correlation scale.} We also considered this hypothesis to be unsatisfactory,
as there are no known examples of an isolated AGN with strong, broad
Balmer lines, but no corresponding narrow lines at all.  A more intriguing
possibility is that the b-system is a black hole ejected from the nucleus
of a galaxy now just hosting the r-system; however, the large
implied ejection velocity
and the light source for the absorption system are problematic.

These conclusions are challenged by \cite{wl}, however,
who obtained VLA observations
of J1536+0441, showing it to comprise two sources separated by $0\asec97,$
with $1.17\pm0.04$ and $0.27\pm0.02$ mJy flux strengths or a 4.3
flux ratio at 8.5 Ghz.  If these separately correspond to the
two broad line systems, then this would be a strong refutation of
hypothesis that J1536+0441 is a strongly bound binary system.
Prior to publication
of the \citet{wl} observations, we requested a single
orbit of {\it HST} imaging to check for multiple sources or unusual
morphological features that might be diagnostic.  We also
obtained long-slit spectroscopy.  These new observations appear
to rule out the ``chance superposition'' or ``ejected black hole''
hypotheses, but leave the choice between the ``double-peaked''
or ``binary black hole'' hypotheses unresolved.

\section{The {\it HST} Imaging Program}

\subsection{The Observations and Their Reduction}

{\it HST} images of J1536+0441 were obtained on April 21, 2009 under program
GO/DD 11993.  The QSO was centered in the PC1 CCD of WFPC2.  The duration of the
program was a single orbit, sufficient to probe the morphology of the
bright AGN emission and to identify bright companions.  Eight short (80s)
exposures were obtained in filter F675W.  The bandpass of the filter includes
the H$\beta$ lines of both the b and r redshift systems, as well as
the [O III] lines associated with the red system.  The F675W images were
dithered to provide $0.5\times0.5$ pixel subsampling to produce a
Nyquist-sampled image of the system, as well as to recover the best spatial
resolution available for the particular camera and filter combination.  Two
exposures were obtained at each dither position to permit the repair of
cosmic-ray hits.  The typical peak exposure in the center of the QSO
subimages was $\sim10^4~e^-.$
The time remaining in the orbit after completion of the dither sequence
was used to obtain two 100s spatially-coincident exposures in the
F439W filter to provide some level of color information,
but these exposures were unfortunately much shallower,
having a peak level of $\sim1.6\times10^3~e^-.$

An up-sampled F675W superimage was generated from the eight individual
exposures.
The two exposures at each dither point were first combined, with cosmic rays
identified as statistical outliers in each pair.
The four summed exposures at each dither position were then combined
to produce a single image with a $2\times$ finer pixel scale
($0\asec0228$/pixel), using the Fourier technique of \citet{l99}.
For the F675W bandpass, the {\it HST} optical spatial band-limit
falls well short of the Nyquist scale in the up-sampled image, thus
high spatial frequencies between the two scales only encodes noise
and were hence eliminated from the final image by a \citet{wiener} filter.
The final F439W image is just an addition of the two sub-exposures after
cosmic ray rejection.  Both images are shown in Figure \ref{fig:image}.

\subsection{The Companion Object}

The F675W image of J1536+0441 shows the QSO to have
a faint spatially-resolved companion.  Its
projected separation from the QSO is $0\asec96$ or 5.0 kpc
in the r-system frame ($z=0.388$) at $90\sdeg1$ position angle.
\footnote{We assume $H_0=71~{\rm km~s^{-1}~Mpc^{-1}},$ $\Omega_m=0.27,$
and a flat cosmology throughout the paper.}
This location is well within the $0\asec03$ position
error of the companion radio source discovered by \citet{wl}.
As we were completing this manuscript, \citet{dec} reported
seeing this companion in VLT K-band images.

We deconvolved the F675W image using 20 iterations of the Lucy-Richardson
algorithm \citep{lucy, rich} to gain a model-independent understanding
of the morphology of the companion.  The PSF was constructed from standard
stars observed over the last several years for photometric
monitoring of the F675W filter.  While, no F675W PSF observations were
available close in time to the present observations, the attendant small
uncertainties in the PSF do not affect our results.
The F675W apparent magnitude of the object
is 20.67 (AB) in a $0\asec5$ aperture.  At $z=0.388,$ the F675W bandpass is
very nearly centered on the SDSS {\it g-}band.  The implied
absolute magnitude is $-21.41,$ which includes
the k-correction and a 0.13 mag correction for galactic extinction.
The companion is unfortunately not visible
in the F439W image; however, the zeropoint in this image is 2.85 mag (AB)
brighter than that for the F675W superimage.  Even if the companion object
had a flat spectral energy distribution, the expected peak count in the
F439W images would only be $\sim0.5$ ADU.  Since the
F439W filter falls to the blue of the 4000\AA\ break at the QSO redshift,
it is likely that the true flux of the companion is even fainter.
The color constraints offered by the F439W images are thus
only of marginal interest.

\begin{figure}
\plotone{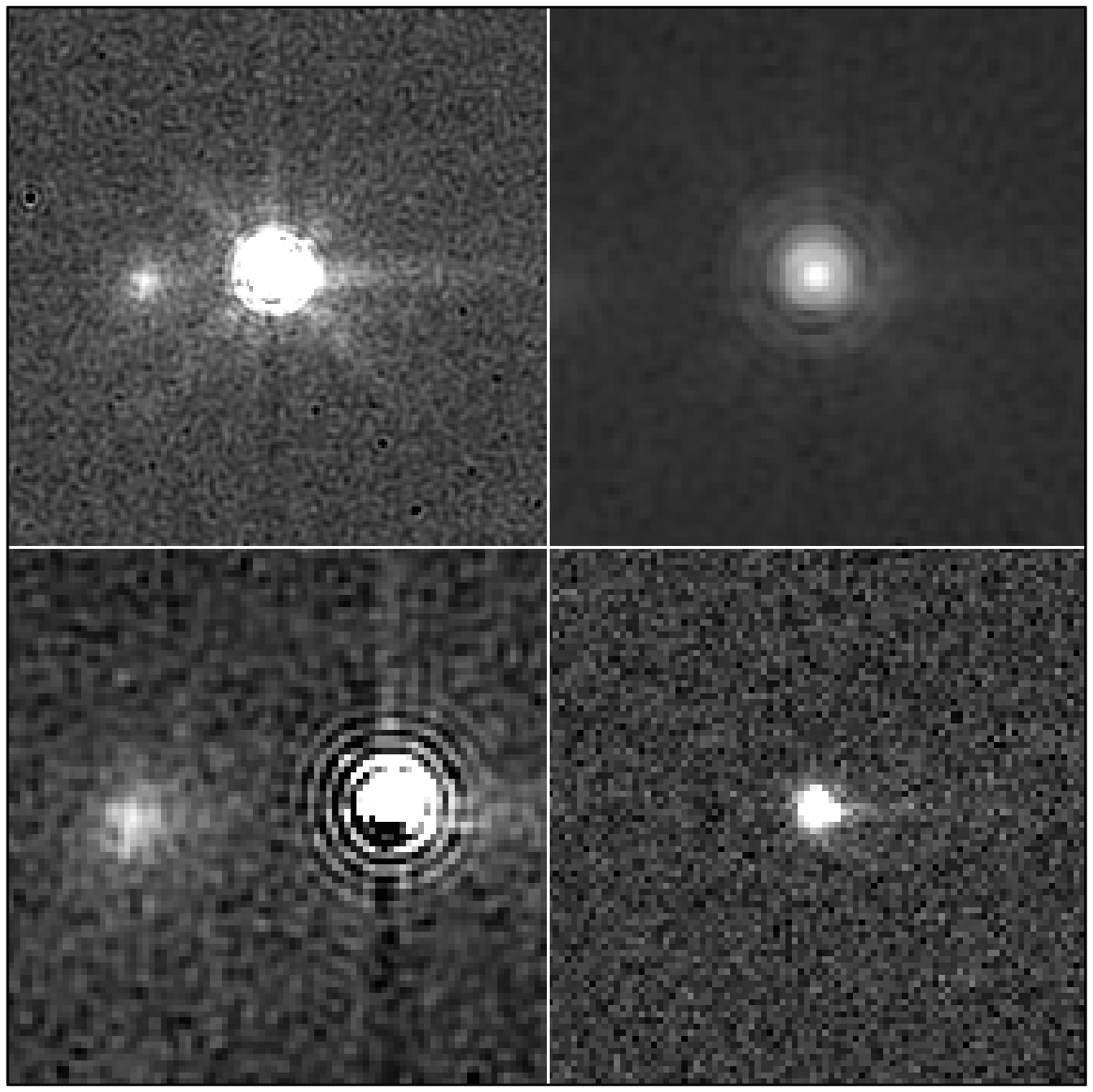}
\caption{{\it HST} WFPC2/PC images of J1536+0441.  North is at the top.
The upper-left panel shows the central $4''\times4''$ of
the F675W up-sampled summed image.
A linear stretch has been used.  The companion is left of the QSO.
The upper-right panel is the central
$2''\times2''$ of same image, but with a
logarithmic stretch to show the inner structure of the QSO core.
The lower-left panel shows a $2''\times2''$ patch of the F675W image
centered between the QSO and companion after a PSF scaled to the peak
of the QSO has been subtracted.
The bright residuals are artifactual.
The lower-right panel shows the central $4''\times4''$ of the
F439W image.  The stretch is identical to that in the upper-left panel.}
\label{fig:image}
\end{figure}

The companion is best seen in the lower-left panel of Figure 
\ref{fig:image}, in which we have attempted to subtract the QSO,
which was done by scaling the reference F675W PSF to the peak flux
of the QSO.  The companion appears to be smooth and slightly elliptical
in morphology.  No tidal features are seen, nor is there any apparent
connection to the QSO.
However, the image is very shallow and we estimate that it would
be difficult to see isolated features with surface brightnesses dimmer
than $\sim20.5,$ thus
this constrains only very strong interactions.

\begin{figure}
\plotone{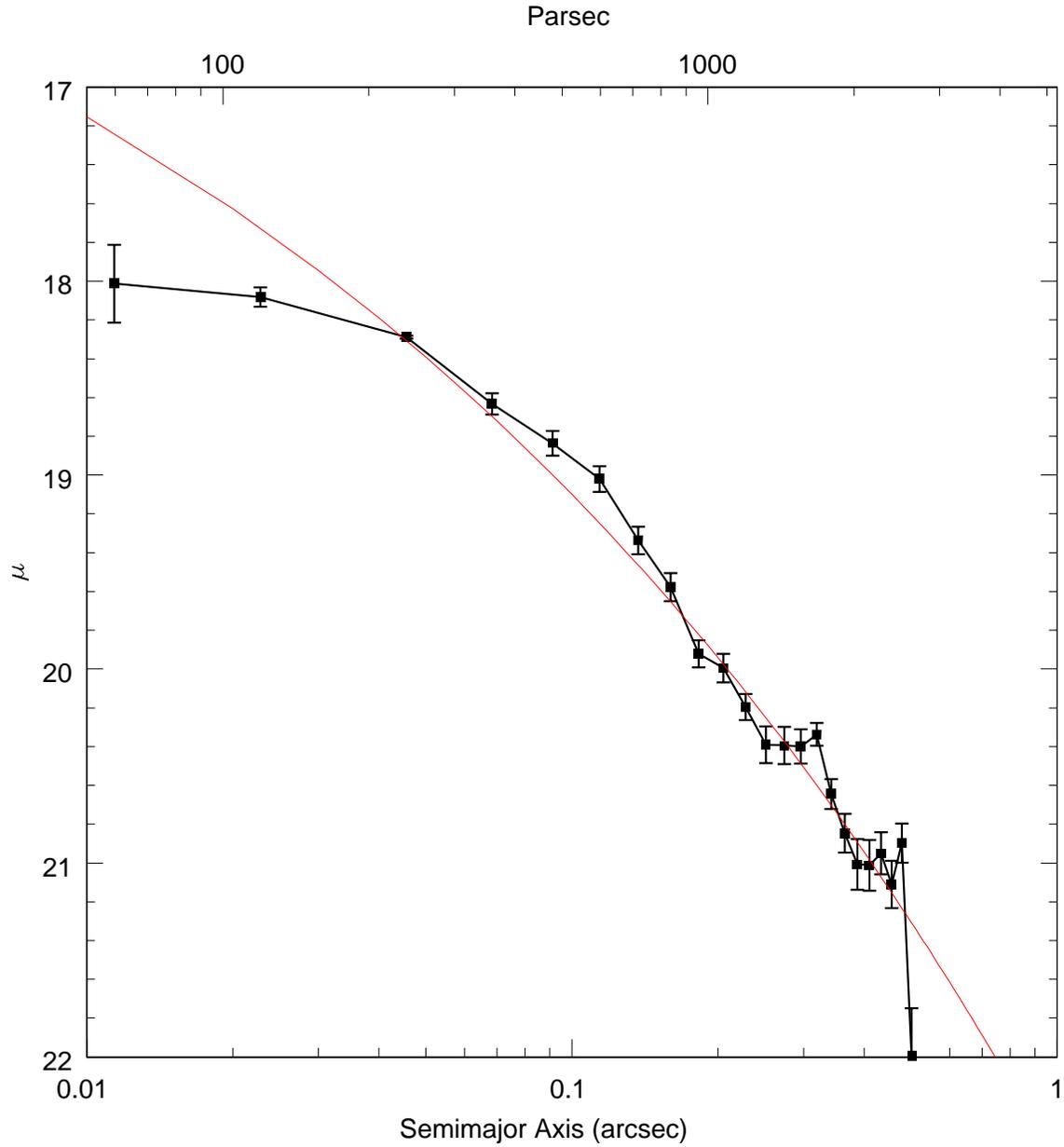}
\caption{The deconvolved brightness profile of the companion
object is plotted (black).  The error bars are based on scatter
observed around the individual isophotes; and
uncertainty in the deconvolution for the central point.
The surface brightness is as observed ---
no extinction or cosmological corrections have been applied.  The
bandpass at the QSO redshift is essentially SDSS {\it g-}band.  The
red line shows an $r^{1/4}$ law fitted to the profile for $r>0\asec04.$}
\label{fig:prof}
\end{figure}

The brightness profile
is presented in Figure \ref{fig:prof}.  Its form resembles that
of an elliptical galaxy with a core \citep{l95}. In contrast, the
profile is less well described by an exponential;
if an exponential is fitted to just the outer part of the profile,
the center of the profile rises well above it.
An $r^{1/4}$ law fitted to the profile yields an
effective radius, $R_e=1\asec23,$ or 6.4 kpc.
If used to estimate a total apparent luminosity,
this form yields a value substantially brighter than the
aperture magnitude quoted above; however, given that the brightness
profile cannot be detected at $r>0\asec5,$ still a small fraction
of $R_e,$ it is difficult to credit such a large extrapolation.
In any case, the value of $R_e,$ as well as the surface
brightness as a function of physical radius, is consistent with that
seen in early-type galaxies within a generous range of
absolute luminosity about the aperture magnitude quoted above.

The most interesting aspect of the companion is its relatively strong
radio emission. \citet{wl} derive $\nu L_\nu$ radio luminosities at 8.5 Ghz
of $5.2\times10^{40}$ ergs s$^{-1}$ and $1.2\times10^{40}$ ergs s$^{-1}$
for the QSO and companion. respectively.
The QSO/companion radio flux ratio is thus only 4.3, as compared
to the optical flux ratio of $\sim30$ implied by the F675W 16.99 (AB)
apparent magnitude of the QSO.
At the same time, \citet{dec} find the K-band flux ratio to be $\sim5,$
which would imply that the companion object is very red, with the
caveat raised above that our small aperture may have underestimated
the total optical luminosity of the companion.
The companion radio luminosity is consistent with the
largest values appropriate to the most luminous ``normal'' early-type
galaxies \citep{sad, wh} --- it is well above the typical power
seen in galaxies close to the estimated absolute magnitude given above.

Despite its strong radio emission, the companion shows no obvious nuclear
point-source component in the optical.
This is consistent with the analysis shown in the next section that
shows that the companion is not the source of the optical emission lines.
\citet{dec} assert that the companion does have a point source component
in contradiction of the present results.  However, they present no analysis or
figures to support this claim, and the spatial resolution of their K-band image
appears to be at least an order of magnitude poorer than
that of the {\it HST} images.

There are no other obvious sources associated with
either the QSO or companion.   The reference PSF and QSO core have identical
morphologies within their half-light radii.  In particular, we can easily rule
out the presence of an additional point source having a flux exceeding 10\% of
the QSO centered at any distance greater than $0\asec1$ from the QSO. 
At slightly larger radii, the
residuals of the PSF subtraction show a largely dipole pattern,
with excess light seen to the NW of the QSO in Figure \ref{fig:image}.
While, as noted earlier, there is very little recent stellar imaging available
in the F675W filter, examination of standard stars observed in the F555W or
F547M filters a few months prior to execution of the present program show
this pattern to be common; it is consistent with a small amount of comatic
aberration.  We thus conclude that the residual pattern is artifactual.
The residuals also make it
difficult to set any constraints on the properties of the QSO host galaxy.
On the assumption that the QSO subtraction residuals are largely due
to a host galaxy, rather than a miss-match between the PSF and QSO core,
the implied host luminosity is less luminous than $M_g\sim-22.$

\section{The Spectroscopy}

\subsection{Observations and Their Reduction}

Spectra of J1536+041 were obtained on the night of April 22 (UT) with the
RC spectrograph on the Mayall 4m telescope at Kitt Peak.  The BL 420 grating
provides a 1.5~\AA\ pixel$^{-1}$ dispersion; in combination
with a $1\asec5$ wide slit, the resulting resolution is
between 3.0 and 3.6~\AA\ over the wavelength ranges covered.
Two grating tilts were used, a blue setting to cover the region from
5000~\AA\ to 7600~\AA\ and a red setting to cover the region
from 7000~\AA\ to 9500~\AA.
The slit was $49''$ long, and was oriented east-west to capture
the companion seen in the {\it HST} images.
The spatial pixel scale is $0\asec69$ per pixel.  Five exposures
were obtained with the blue setting, and four exposures were obtained with the
red setting.  All exposures were 900 seconds.  The seeing was 
poor during the period of the observations;
a gaussian fit to the object flux along
the slit gives about $1\asec7$ FWHM. In addition to the exposures on the
object itself, helium-neon-argon comparison spectra were obtained at the object
position and flux standard stars were observed with the same setup.

\begin{figure}
\plotone{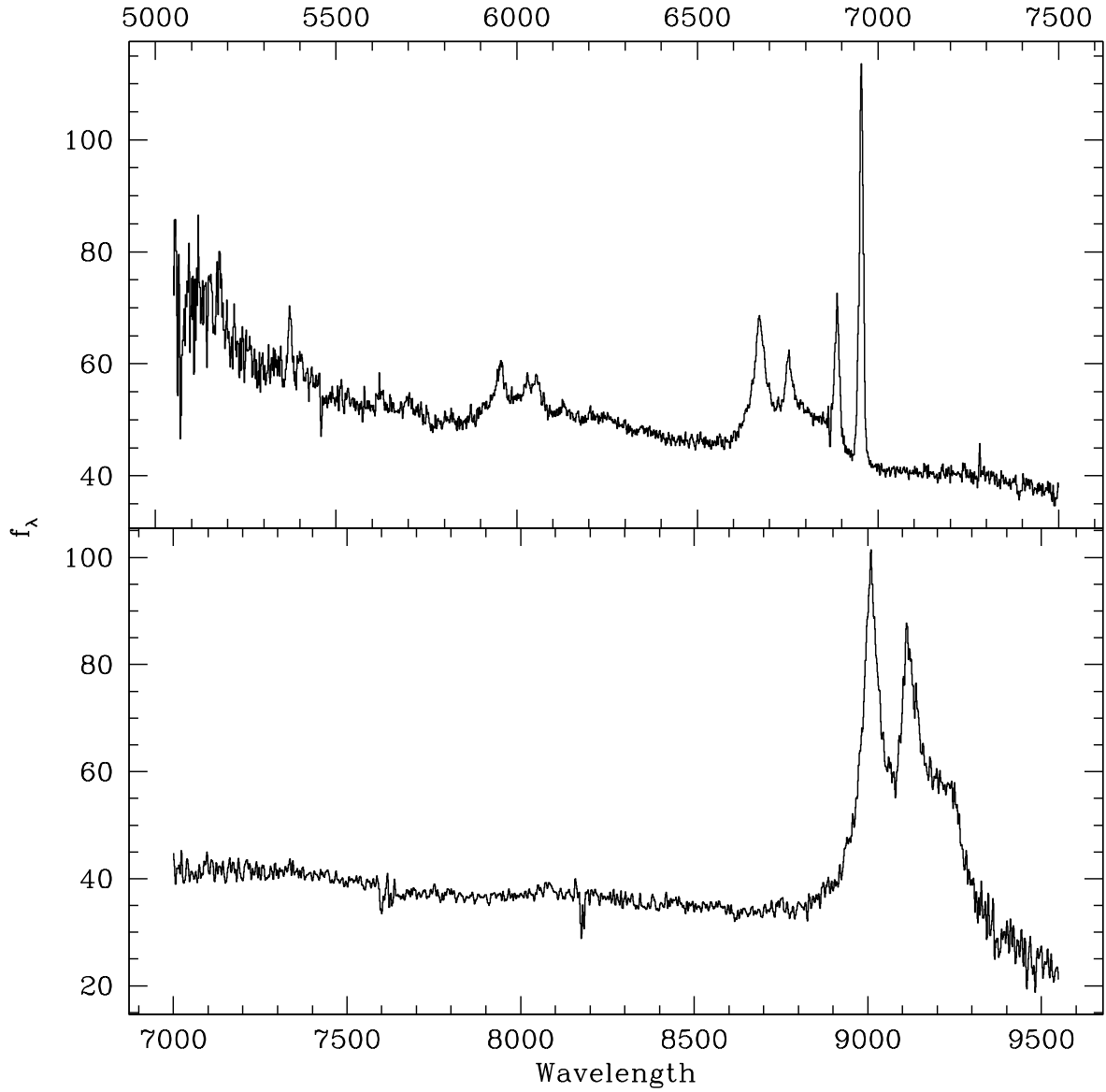}
\caption{The new spectrum of J1536+0441 obtained with the RC
Spectrograph on the Mayall Telescope.
The observed wavelength scale for the bluer region (top panel)
is given on the top of the figure; that for the redder region
(bottom panel) is given on the bottom.  Fluxes are in units of
$10^{-17}$ ergs cm$^{-2}$ s$^{-1}$ \AA$^{-1}$.}
\label{fig:spectra}
\end{figure}

The spectra were reduced by removal of bias, division by a flat field,
subtraction of the sky spectrum, and extraction of a one-dimensional
spectrum from a region of the frame extending about $2''$ on either
side of the spatial peak.  The flux standards were used for both
removal of atmospheric absorption features and for flux calibration.
Positions of several dozen lines in the helium-neon-argon spectra
were fit to low-order polynomials,
resulting in wavelength calibrations accurate to
0.08\AA\ RMS for the bluer setting and 0.09\AA\ RMS for the redder setting.
The individual spectra of each wavelength region were then combined,
though some of the measurements described below were made from the
individual exposures in order to estimate the uncertainty from the scatter.
The final combined spectrum is shown in Figure \ref{fig:spectra}.

The new spectrum is quite similar to the SDSS spectrum.  In addition to the
complicated structure of the Balmer lines, discussed in the next section, the
same narrow lines are seen, including [O II] $\lambda$3727,
[O III] $\lambda$4363, $\lambda$4959, and $\lambda$5007,
and [Ne III] $\lambda$3869 and $\lambda$3968.  In addition, the [N II]
$\lambda$6584 line is now visible on the red side of the r-system
H$\alpha$ line.  The narrow absorption lines of Na I D and Ca II H
are also seen.

Our new spectrum has somewhat higher signal-to-noise
than does the SDSS spectrum in the region between 5500~\AA\ and 6500~\AA\
and the H$\delta$ lines for the b- and r-systems are visible.  In
addition, the Fe II emission between H$\gamma$ and H$\beta$ is easier to
distinguish than in the SDSS spectrum.  We attempted to determine whether it
comes predominantly from the b- or r-system by cross-correlation with the
corresponding region from the Fe II template developed by \citet{bg92} from
the spectrum of I Zwicky 1.  This cross-correlation shows that
the optical Fe II is
primarily associated with the b-system velocity, though the signal-to-noise is
not sufficient to rule out a contribution from the r-system as well.

\subsection{The Balmer Line Profiles}

The new data clearly shows that the combined profiles of the Balmer
lines extend as far to the red from the narrow-line redshift
as they do to the blue, as pointed out by \citet{chorn}.
This is particularly clear in the case of H$\alpha$,
which appears to extend to the red from the r-system at
a constant level, and then drops off at about 9260\AA,
which corresponds to about 4700 km s$^{-1}$
in the rest frame of the r-system.
This makes the bottom part of the profile approximately
symmetric on the red and blue sides.

It is unclear how to measure the strengths, widths, and shapes of the various
components that make up this profile.  Relative to the continuum level,
the b-system peak of H$\alpha$ is 1.24 times the height of the r-system peak.
Relative to the level of the flat part of the red shoulder,
that factor becomes 1.45.  We measure the equivalent
width of the entire H$\alpha$ complex as 348\AA,
with 247\AA\ of that total in the broad base,
defined with a flat top at the level of the red shoulder,
and with 44\AA\ and 57\AA\
as the contributions of the r-system and b-system peaks respectively.

\subsection{Profile Changes between the two epochs}

The most compelling confirmation that this system is a supermassive black hole
binary would be the measurement of changes in the radial velocities
of the two peaks consistent with orbital motion.
In \citet{tnt} we estimated the separation of the two black holes
to be 0.1pc with a $\sim100$ year orbital period, based solely
on the implied radial velocity difference, an estimate of the
two black hole masses, and the assumption of mean values for
the orbital phase and inclination.
These numbers predict a measurable change in velocity separation
of the two peaks in as little as a year.  Our new spectrum
was obtained 1.04 years after the SDSS April 2008 observation,
which corresponds to 0.75 years in the rest frame of the object.

Visual inspection of the new spectra,
as well as the report of \citet{chorn}, shows
that changes in the separation of the peaks of the Balmer
lines are small or absent.
Until and unless these changes become significant,
improved constraints
on the orbital parameters are based instead on the size of the errors
bounding a null measurement.

Given the complex structure of the Balmer-line profiles,
measurement of accurate velocity requires that the spectra at all
epochs and from all sources be treated consistently.  In particular,
a consequence of the composite nature of the profile is the
possibility that the H$\alpha$ and H$\beta$ lines in each system may
have different behavior.  We note specifically that there is a 
well-established trend of decreasing Balmer decrement with velocity
away from the line center in Seyfert galaxies and QSOs \citep{shud, crenshaw}.
That is, the wings of the H$\alpha$ line fall off more quickly than those
of the H$\beta$ line.   
This effect appears to be present in the r-system Balmer lines,
and it results in a larger gradient under the b-system H$\alpha$
line than under the H$\beta$ line, leading to the perception that
the splitting between the two systems is less in the H$\alpha$ line than
in the H$\beta$ line.  However, if one overplots the
H$\beta$ and H$\alpha$ regions with a relative shift equal to the
ratio of the rest wavelengths of the lines, it can be seen that the
peaks of the two b-system lines are in excellent velocity agreement.

In order to minimize such effects, we have remeasured the
SDSS spectrum and our new spectrum in the same way,
emphasizing the peaks of the lines, which should be less
sensitive to underlying gradients.  We find that relative to the
[O III] lines, the r-system H$\beta$ peak has a velocity of
$61\pm15$ km s$^{-1}$, while the b-system H$\beta$ peak has a velocity
of $-3484\pm35$ km s$^{-1}$.  The uncertainties quoted are
the mean error as derived from the scatter in the five independent
observations.  The corresponding velocities from the
SDSS spectrum are 31 km s$^{-1}$ and $-3512$ km s$^{-1}$
for the r- and b-system H$\beta$ peaks respectively.
Thus, the H$\beta$ splitting has changed from $-3543$
to $-3545\pm38$ km s$^{-1}$.  Similarly, we measure the
H$\alpha$ splitting as $-3514$ km s$^{-1}$ in the SDSS
spectrum and $-3497\pm40$ km s$^{-1}$ in our new spectrum.
A final caveat is that there is some Balmer emission
associated with the narrow lines though it is difficult to isolate, 
and this may influence
the measured velocities of the r-system broad lines.

In summary, we see no shift in the positions of the Balmer line
peaks over this period, with a two sigma limit of about 80 km s$^{-1}.$

Lastly, we investigate changes in the strengths of the peaks using
the H$\beta$ line because it is well covered in both spectra.
In addition, the [O III] lines should
not change and so they provide some check on our result.  After rebinning the
new and SDSS spectra to the same resolution, we find that changes in the strengths
of the two peaks relative to the continuum and to each other can be limited to less than 
3\%\ over the period between the two observations.

\subsection{Spatial Location of Spectral Features}

\begin{figure}
\plotone{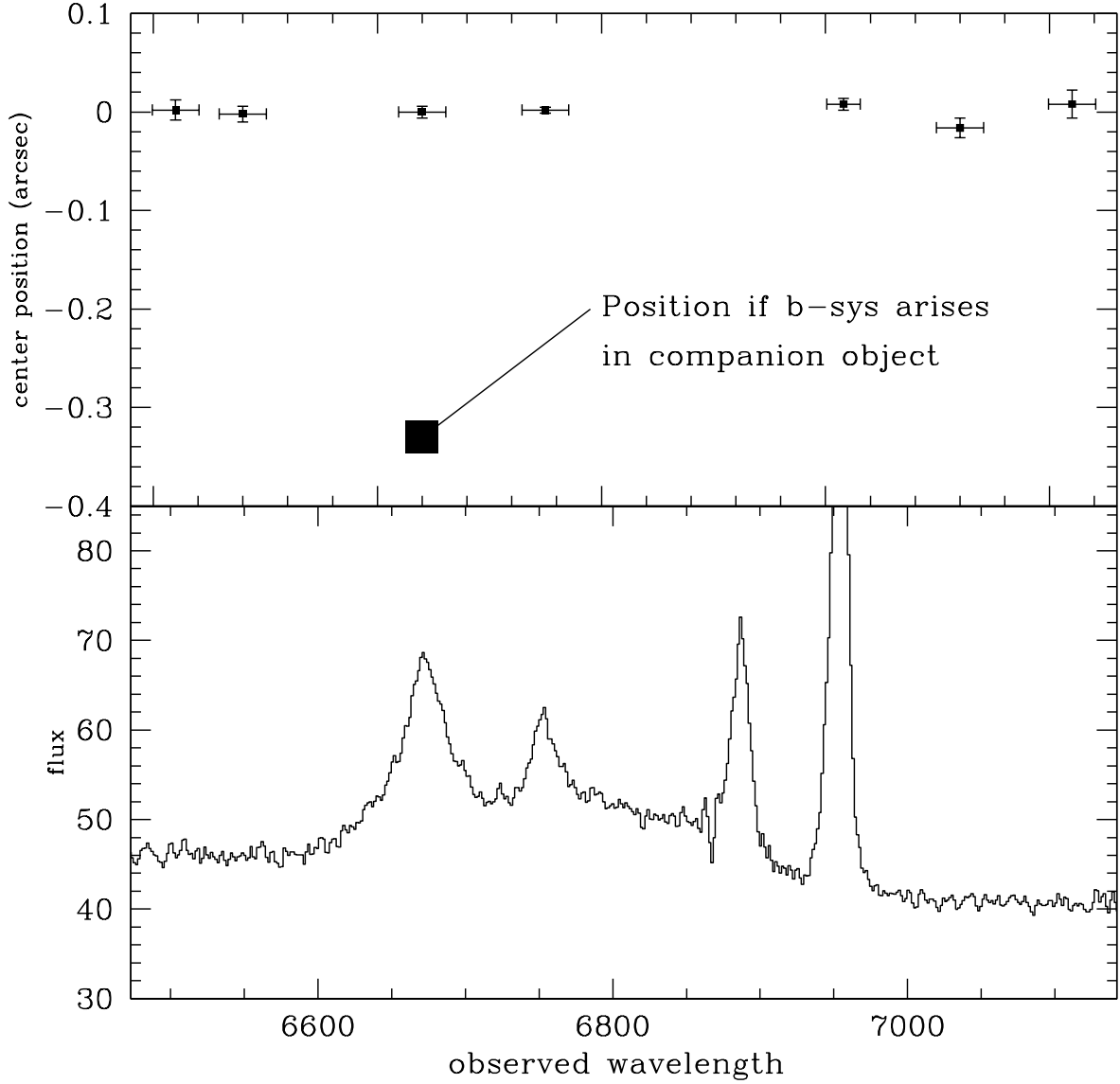}
\caption{Top panel shows the spatial location along the slit of the
flux-weighted centroid of the spectrum in bands shown by the horizontal
error bars.  The vertical error bars show the standard deviation in
the measured positions from five independent exposures.  The lower panel
shows the spectral flux through this region.  The bands have been chosen
to represent (from left to right) two continuum regions,
the b-system H$\beta$ line, the r-system H$\beta$ line,
the [O III] $\lambda$5007 line, and two more continuum regions.  The
square indicates where the point representing the b-system
H$\beta$ line would fall if it arose in the companion object.}
\label{fig:hbshift}
\end{figure}

\citet{wl} speculate that one of the two strong peaks in the Balmer line
profiles may arise in the companion source seen in the radio.
Since our slit was aligned
east-west, and the separation between the two objects is approximately
one arcsecond, this would result in an easily visible offset in the
spatial location of the spectrum at the position of any feature arising
in the companion.  To test this idea, we fitted the intensity of the
spectrum along the slit as a function of wavelength at positions
corresponding to the b- and r-system H$\beta$ lines, the two [O III] lines,
and four continuum points spanning this region.  The centers of the
best-fit gaussians are shown in Figure \ref{fig:hbshift}.
We have removed a linear tilt in the spectrum, due to a slit
misalignment of the dispersion with respect to the CCD rows.
The horizontal error
bars show the range of the columns of the frame averaged.
The vertical error bars show the standard deviations in the five
measurements made from the five independent exposures.
The corresponding spectrum is
shown in the lower panel of Figure \ref{fig:hbshift}

An estimate of the expected shift if the b-system arose in
the companion object can be obtained by noting that
the equivalent width of the b-system H$\beta$ line is approximately
15\AA\ in the ``composite'' spectrum and that
the width of the region averaged together is approximately 30\AA.
Since the ratio of flux in the companion object is only about
1/30 of that in the primary object within this spectral region, not much of the
continuum can arise in the companion.
This results in a relative flux within this 30\AA\ window
of about 2 to 1 in favor of the brighter object, producing
a position that would be offset by 1/3 of the one arcsecond separation.
This is obviously incompatible with the measured positions,
leading us to conclude that all of the spectral features arise
in the primary object.

\section{What J1536+0441 Could Be and What It Isn't}

The analysis of the new {\it HST} images and KPNO spectra of
J1536+0441 appears to rule out some hypotheses on its physical
configuration, while leaving the door open to others that are likely
to require additional observations for further evaluation.
We close by summarizing what we conclude is the current
status of the various explanations for the unusual properties of J1536+0441.

\subsection{Chance Superposition of Objects}

The discovery of \citet{wl} that J1536+0441 is a well-resolved binary
radio source appears to argue for the conclusion that it is simply the
line of site superposition of two AGN,
a conclusion also advocated by \citet{dec}.
\cite{tnt} did not favor this hypothesis, as
the probability of a superposition was low ($3\times10^{-3}$),
and required one of the AGN to have unique behavior in
exhibiting broad lines without any corresponding narrow lines.
In advance of the {\it HST} observations, we might have expected to see
the QSO resolved into two point sources, corresponding to two physically
separate AGN.  Alternatively, we might have seen an optical jet emerging
from the QSO.  The optical counterpart to the less-luminous radio
source, however, appears to be a normal galaxy with no evidence for
an optically visible AGN, based both on the {\it HST} imagery, and analysis
of the angular centroids of the emission lines in the long-slit spectroscopy.
Superficially, it thus appears that this object is merely an ``innocent
bystander'' to J1536+0441, although its large radio luminosity
remains unexplained in this scenario.
One possibility, however, is that it
at least is the source of the absorption lines seen in its spectrum.
The ``a-system" has a projected velocity of $240~{\rm km~s}^{-1}$
less than the r-system, compatible with galaxy and the QSO forming
a binary system.

\citet{wl} derived a
considerably higher {\it a priori} probability of superposition
than our estimate quoted above.
Using the \citet{h06} QSO-QSO correlation function,
they derived a probability of about unity
for the occurrence of a 5 kpc separation pair within
the size the original sample studied by \citet{tnt}.
We note, however, that \citet{h06} limited their sample of QSO pairs
to those with $\Delta V<2000~{\rm km~s}^{-1},$ on the assumption
that larger velocity separations would not be expected in systems
that are truly physically associated.
\citet{croom}, for example, show no significant correlation
between QSOs with velocity differences as large as that in J1536+0441.
A possible exception to this
would be for two QSOs both bound in the potential of an extremely rich
cluster.  \citet{heck}, for example, hypothesize that the
two emission line systems separated by $2650~{\rm km~s}^{-1},$ in the QSO
SDSSJ092712.65+294344.0 is a system analogous to the
$\sim 3000~{\rm km~s}^{-1}$ difference between the two emission
line systems seen in NGC 1275 \citep{min}, the first-ranked galaxy of the
Perseus galaxy cluster.  The probability of this occurrence for
J1536+0441 should be encoded in the \citet{croom} function, however.

The limit on spatial coincidence from the new {\it HST}
images, however, is now at least two orders of magnitude more stringent
than that assumed in \citet{tnt} or \citet{wl}.
This argument continues to disfavor the idea that there are two
separate host galaxies.

\subsection{Ejected Black-Hole}

If one assumes that the r-system, which includes the narrow line system,
represents the rest velocity of the host galaxy, then in
advance of the present observations, one might have hypothesized
that the b-system represented a black hole ejected from the nucleus.
There are two ways to achieve this result.
Under the first mechanism, as two black holes
merge, an asymmetric jet of gravitational radiation can be emitted, propelling
the merged black hole out of the nucleus.  In the second mechanism an infalling
black hole interacts with pre-existing binary black hole, and is ejected
in an exchange of orbital energy that binds the binary tighter and
expels the third black hole.

In the case of ejection of a merged black hole, the 3500 km s$^{-1}$
velocity of the b-system is at the edge of plausible ejection velocities
that have been demonstrated in numerical experiments \citep{camp,gonz}.
Such high velocities depend on very specific mass ratios, spins, and
orbital parameters of the merging black holes, and appear to be extremely
unlikely. \citet{schnitt} show that the probability of ejection velocities in
excess of $\sim3000$ km s$^{-1}$ may occur in $<<0.1\%$\ of all black hole
mergers.  Regardless, while this might
explain the QSO as two separate objects, one of which has been ejected, the
spatial centroid of the H$\beta$ lines show that neither of these objects resides
in the companion galaxy.  

Ejection as the result of the interaction of an infalling black hole
with a pre-existing binary black hole may also produce relative
velocities in excess of $\sim3000$ km s$^{-1}$ under rare conditions.
The distribution of ejection velocities presented by \citet{hoffman}
show that these may occur in a few percent of the interactions, thus
apparently making this mechanism much more likely to produce a
high-velocity ejection than the merger case.  This mechanism allows
for the existence of two AGN after the ejection, and thus might have
explained J1536+0441 had the present observations revealed two distinct
sources of optical emission.  As with the merger case, however,
it cannot explain the spectrum of J1536+0441 as a single source.

On observational grounds, we did not favor this hypothesis in \citet{tnt},
as the high velocity of the b-system, but the unresolved nature of the
point source implied either that the system was being observed close
to the epoch of its creation, or that the ejected black hole was essentially
traveling along the line-of-sight.  It was also difficult to understand
the geometry under which the b-system AGN could serve as a continuum source
against which the a-system absorption lines are seen.

If the a-system is attributed to the companion galaxy, then the geometry
under which the a-system is seen against the QSO may be more plausible.
The tighter restrictions placed on the ejection geometry by the {\it HST}
imagery makes this even less attractive, however.

We also note the superficial similarity of J1536+0441 to the object
HE 0450-2958, which has been proposed as a ``naked''
QSO \citep{magain05,khpi}, possibly ejected 
from a disturbed galaxy about $1\asec5$ away \citep{lmcl}.
Although the projected separation between the QSO and the putative host,
6.5 kpc, is similar to the separation between
J1536+0441 and the companion object, there are two critical differences.
First, the upper limit we have derived for an underlying host
galaxy at the position of the J1536+0441 nucleus, $M_g\sim-22$,
is still bright enough that no alternative host is required.
Second, this explanation does not account for the unique spectrum of 
J 1536+0441; the QSO in the GE 0450-2958 system has a normal spectrum.

\subsection{Double-Peaked Emitter}

The most attractive explanation for J1536+0441 may indeed be that it
is a ``double-peaked'' emission line QSOs.  The red bumps
in the broad lines discovered by \citet{chorn}, and confirmed here,
certainly resemble some of the broad and diffuse emission seen in these
objects.  The double-peaked class is further already known, and admits
a variety of unusual broad line profiles.  At the same time, there are
striking differences between the Balmer line profiles in J1536+0441 and
those of the double-peaked objects generally.

\begin{figure}
\plotone{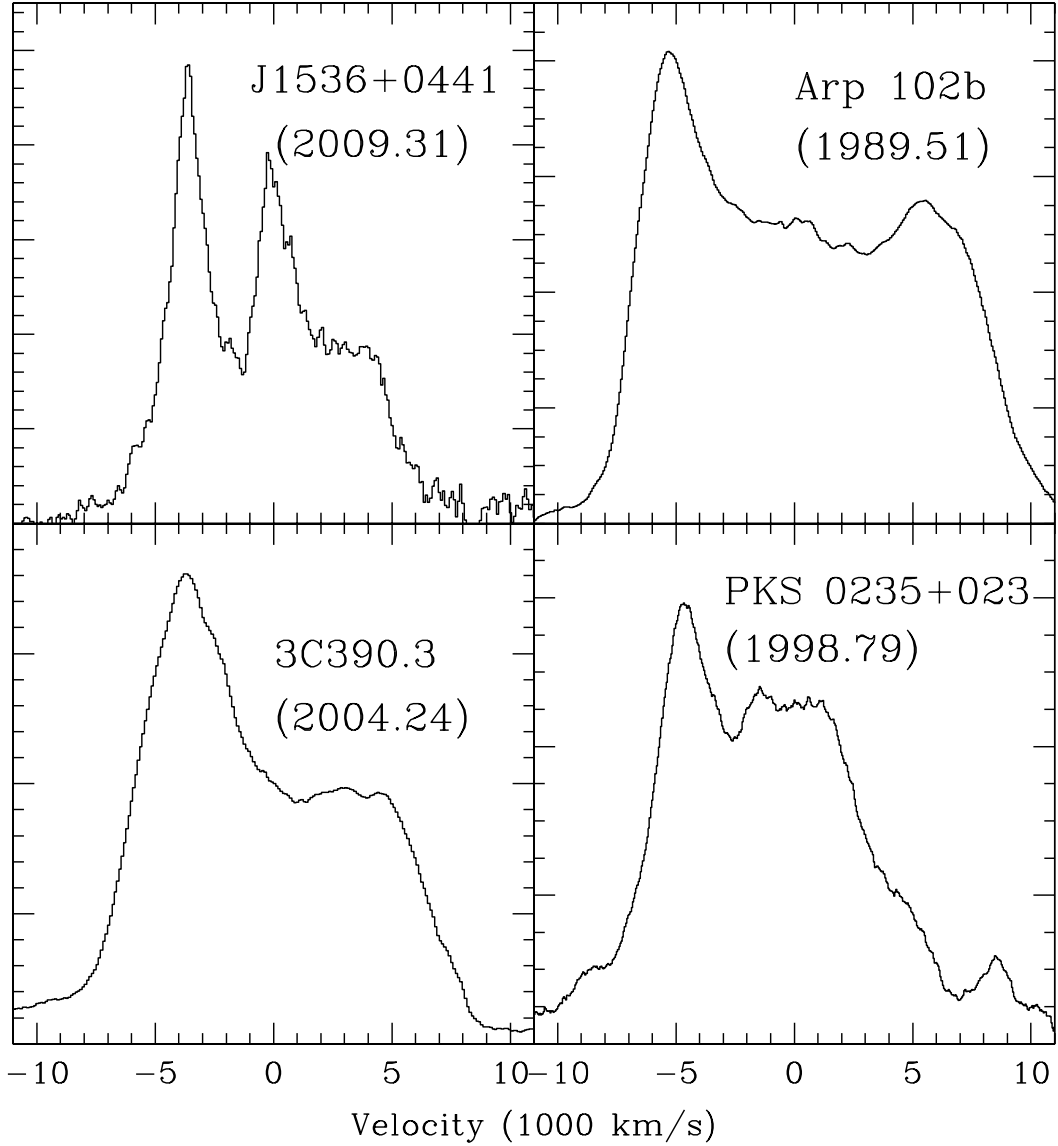}
\caption{Comparison of the H$\alpha$ profile of J1536+0441 with three
objects classified as double-peaked line emitters potentially
analogous to J1536+0441 from the study by
\citet {ghe}.  Each object is identified and the epoch of the observation
is given.  The epochs of the comparison objects were chosen as those
with profiles that most closely matched J1536+0441 in terms of the ratio
of the strengths of the two peaks.  The narrow lines have been removed
from the comparison objects, and the continuum has been subtracted from
all the objects.  The narrow lines in J1536+0441 are insignificant.}
\label{fig:dpecomp}
\end{figure}

The prototypical profile of the double-peaked objects has two rounded peaks
of roughly equal strength joined by a flat or depressed central plateau.
This form can be well fitted by simple relativistic accretion disk models
\citep{chf,strat,ghe}.
These objects occasionally show substantial departures from this form, in
which relatively sharp, transient peaks are seen.
Michael Eracleous at our request selected three
cases of such profiles from the compilation of \citet{ghe};
these are shown, together with the H$\alpha$ profile of J1536+0441
in Figure \ref{fig:dpecomp}.
The narrow lines, H$\alpha$, [N II], [S II], and [O I],
have been removed from the double-peaked lines.
In the case of J1536+0441, the narrow lines are very weak;
the [N II] $\lambda$ 6584 line is just barely visible on the red wing of the
r-system H$\alpha$ line and the other lines are not detectable at all.

Although we admit that the detailed accretion disk structure that leads 
to these profiles is not well understood,
we observe several notable differences between J1536+0441
and these objects.  (1) J1536+0441 has a strong, central, broad peak, which is
not seen in any of the double-peaked emitters.  (2) The blue peak in J1536+0441
is very strong and very narrow at its top, surpassing, though perhaps by only
a small amount, these characteristics in any of the other objects.
(3) The strong, sharp peaks in the double-peaked emitters are transient.
In all three of these objects, the blue peak showed large changes in its
intensity over a few months to less than a year.  J1536+0441,
on the other hand, has shown a constant profile for almost a year now.

It is worth noting that the double-peaked
emitters themselves were initially advanced as binary-black hole candidates.
It is only the failure of the lines to vary in velocity over long time spans
that has led to alternative interpretations of this class.  Nor is
there a clear theory on how the asymmetries accretion disks in these
systems are generated or maintained.  If J1536+0441 is a member of
this class, then it may well motivate an improved picture of how
disk instabilities arise and evolve.

\subsection{A Binary Black Hole}

The new observations presented here and those of \citet{chorn} do
not advance the hypothesis that J1536+0441 hosts a binary black hole,
except by perhaps eliminating the superposition and ejected black hole
hypotheses.  The lack of any observed velocity shift in the b-system
or r-system over the span since the original SDSS spectrum was obtained
provides additional constraints on the orbital parameters.
The prediction that velocity shifts could be seen within a year
was based on the presumption of an average
viewing geometry and orbital phase for the binary system.  If, however,
the velocity difference between the b and r-systems represents a
substantial portion of the orbital velocity, then the orbital
radius of the system is larger, and the period longer than the
estimates given in \citet{tnt}.  The geometry also
requires longer intervals to see velocity changes if the system
is close to quadrature.

The derived orbital parameters also depend on the total mass of the binary
system, which was estimated by \citet{tnt} from the widths of the H$\beta$
lines and the continuum luminosity.  The assumption that the two black holes
were emitting at the same fraction of their Eddington luminosities resulted
in estimates of $10^{7.3}$ and $10^{8.9}$ M$_\odot$ for the b- and r-systems
respectively.  The discovery of  the extension of the Balmer line profiles
to the red allow us to revisit this calculation.  Replacing the former value
of 6000 km s$^{-1}$ with a new measured value of 10,600 km s$^{-1}$ for the
r-system FWHM results in the increase of that black hole to $10^{9.4}$
M$_\odot$.  Two other methods can be used to estimate the black hole masses.
The width of the [O III] $\lambda$5007 line can be used as a surrogate for
the stellar velocity dispersion \citep{ssgb}, yielding a mass of
10$^{8.4}$ M$_\odot$.  Also, one can get a handle on the mass by
adopting an average value of L/L$_{Edd}$, 0.14, for QSOs with the redshift and
luminosity of this object \citep{sgsrs}, which results in a mass of
10$^{8.7}$ M$_\odot$.

\begin{figure}
\plotone{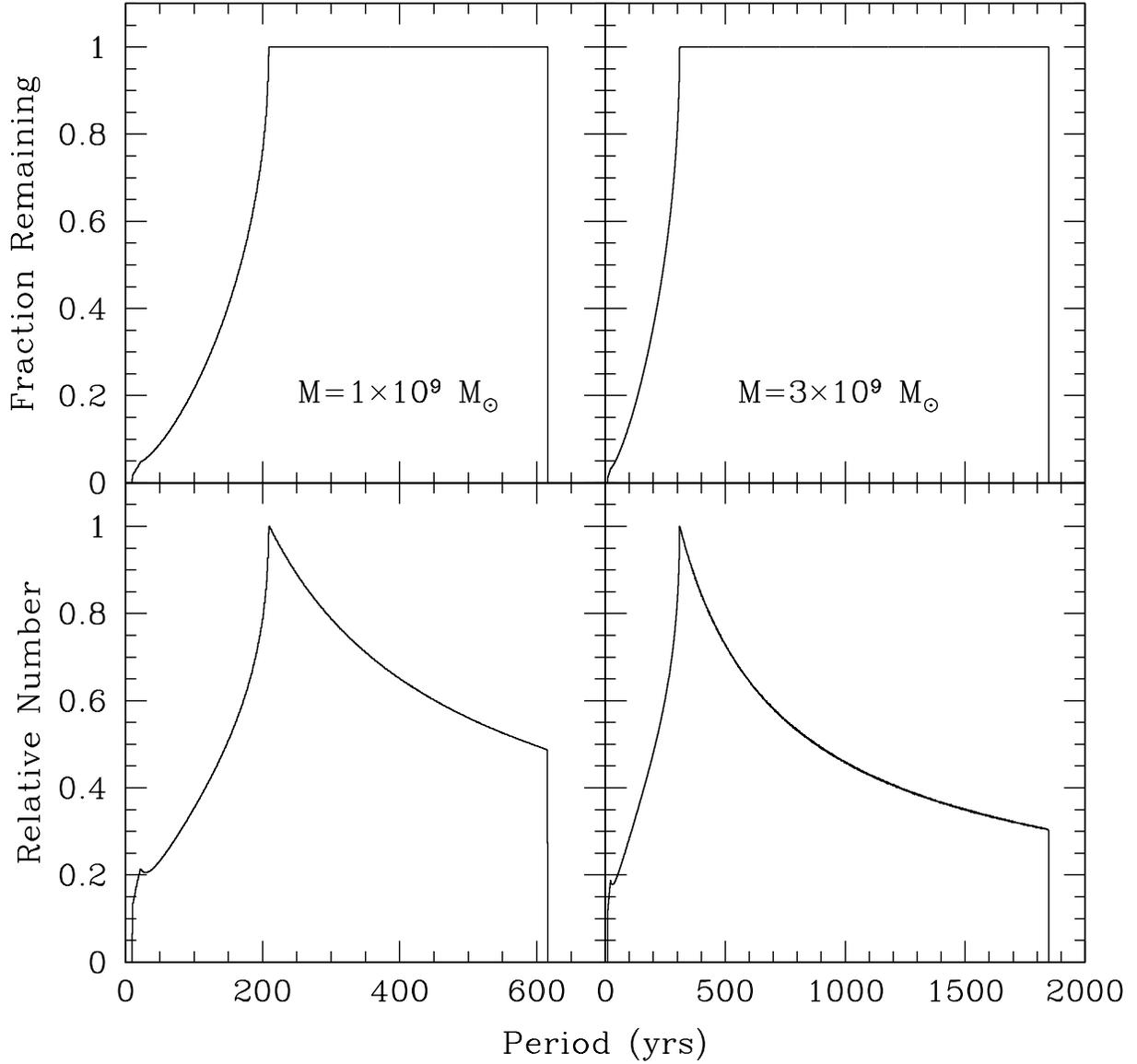}
\caption{Orbital parameter space for binary black hole model,
showing allowed range of inclination and phase at $t=0$ (2008.27).
Excluded regions come from $2\sigma$ limit of 80 km s$^{-1}$ between
original SDSS observation and KPNO observation reported here.
Shown are regions excluded for a total system mass of $1\time 10^9 M_\odot$
(yellow) and $3\times 10^9 M_\odot$ (blue shading).}
\label{fig:phaseinc}
\end{figure}

We explored the new constraints from our spectroscopy in two different ways.  
First, we calculated
the expected velocity change for all possible values of the inclination and
initial phase over the time period between the SDSS spectrum and our
new spectral observation.  We assume that the orbit is circular.
Figure \ref {fig:phaseinc} shows the allowed
and excluded regions, given our $2\sigma$ limit on the velocity change of
80 km s$^{-1}$ for two different values for the total system mass,
$1 \times 10^9$ and $3 \times 10^9$ M$_\odot$.
This confirms the qualitative statement above --- the allowed orbits are those
in which the inclination is closer to
line of sight and the initial phase is closer to maximum elongation.
These are the cases for which the observed velocity difference between
the two systems represents a larger fraction of the true space velocity
difference.

\begin{figure}
\plotone{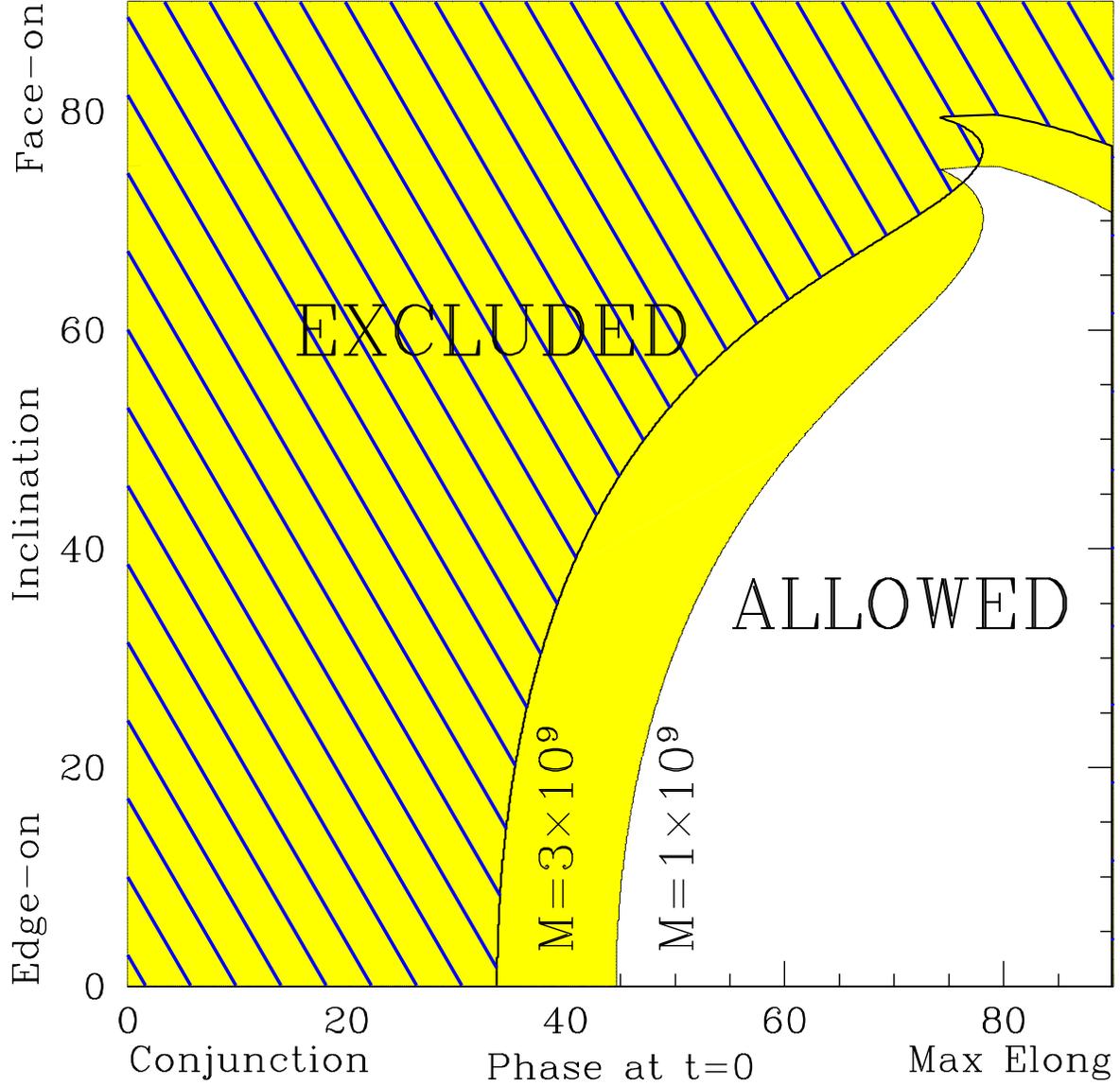}
\caption{The figure shows the fraction of binary orbits as
a function of period that are permitted by
the new spectroscopic observation.  The right and left panels show
calculations for a total system mass of $1 \times 10^9$ and $3 \times 10^9$
solar masses respectively.  The top panels show the fraction of original
possible orbits remaining; the bottom panels show the distribution of those
remaining orbits over period.  The median remaining period is 319 years
for the smaller total mass and 743 years for the larger total mass.}
\label{fig:orbper}
\end{figure}

While this exercise demonstrates the values of initial phase and
inclination that are still permitted,
it does not show the fractional reduction in
allowed orbits, nor does it show the range of allowed periods.  These are
shown in Figure \ref {fig:orbper} for the two system masses we investigated.
The top panels show the fraction of original orbits that remain possible
as a function of the orbital period.  None of the periods longer than 200 years
has been ruled out.  The remaining possible orbits account for 43\%\ and 56\%\
of the original orbits for the smaller and larger system masses.
The lower panels show the relative number of possible
orbits remaining as a function of period.
The median values for the period of the
remaining orbits are 319 and 743 years for the smaller and larger system masses.

As we pointed out, the differences between the observed Balmer line profiles
and those seen in double-peaked emitters, we must also note that the newly
discovered extension of the profiles to the red is problematic for the binary
explanation.  Rather than a model in which two separate massive objects are
orbiting a center of mass located between them, this profile suggests that the
smaller object might be embedded in (or passing through) the disk surrounding
the larger one.  The extension to the red could result from waves or
disturbances in the disk as a result of this interaction.  

It is clear that the ultimate confirmation or rejection of the possibility
that the blue emission-line peak represents an object in orbit around
a central object will come from a number of years of spectroscopic monitoring.

\acknowledgments

We thank Matt Mountain for the grant of STScI Director's Discretion time
under which the {\it HST} observations were obtained.  We thank
the STScI support staff for prompt and expert assistance in the
preparation of the observing program.  We thank Buell Jannuzi for
the availability of the KPNO 4-m time.  We thank Matthew Lallo for
discussions on the properties of the WFPC2/PC1 PSFs.
We thank Michael Eracleous for useful conversations on the properties
of ``double-peaked'' emitters, and for identifying potential
analogues to J1536+0441.  We thank Suvi Gezari for providing us
with the H$\alpha$ spectra of these analogues.

\clearpage

\clearpage

\begin{thebibliography}{}

\bibitem[Boroson \& Green(1992)]{bg92} Boroson, T.~A., \& Green, R.~F.\ 1992,
\apjs, 80, 109
\bibitem[Boroson \& Lauer(2009)]{tnt} Boroson, T.~A., \& Lauer, T.~R.\ 2009,
\nat, 458, 53 
\bibitem[Campanelli et al.(2007)]{camp} Campanelli, M., 
Lousto, C., Zlochower, Y., \& Merritt, D.\ 2007, \apjl, 659, L5 
\bibitem[Chen et al.(1989)]{chf} Chen, K., Halpern, J. P., \& Filippenko, A. V. 1989,
\apj, 339, 742
\bibitem[Chornock et al.(2009)]{chorn} Chornock, R., et al.\ 
2009, The Astronomer's Telegram, 1955, 1
\bibitem[Crenshaw(1986)]{crenshaw} Crenshaw, D.~M.\ 1986, \apjs, 
62, 821 
\bibitem[Croom et al.(2005)]{croom} Croom, S.~M., et al.\ 
2005, \mnras, 356, 415
\bibitem[Decarli et al.(2009)]{dec} Decarli, R., Treves, A., Falomo, R.,
Dotti, M., Colpi, M., Kotilainen, J.~K.\
2009, The Astronomer's Telegram, 2061, 1
\bibitem[Gaskell(2009)]{gas} Gaskell, C.~M.\ 2009, arXiv:0903.4447
\bibitem[Gezari et al.(2007)]{ghe} Gezari, S., Halpern, 
J.~P., \& Eracleous, M.\ 2007, \apjs, 169, 167 
\bibitem[Gonz{\'a}lez et al.(2007)]{gonz} Gonz{\'a}lez, 
J.~A., Hannam, M., Sperhake, U., Br{\"u}gmann, B., 
\& Husa, S.\ 2007, Physical Review Letters, 98, 231101 
\bibitem[Heckman et al.(2009)]{heck} Heckman, T.~M., Krolik, 
J.~H., Moran, S.~M., Schnittman, J., \& Gezari, S.\ 2009, \apj, 695, 363 
\bibitem[Hennawi et al.(2006)]{h06} Hennawi, J. F., et al.\ 2006, \aj, 131, 1
\bibitem[Hoffman \& Loeb(2007)]{hoffman} Hoffman, L., \& Loeb, A.\ 2007,
\mnras, 377, 957 
\bibitem[Kim et al.(2007)]{khpi} Kim, M., Ho, L., Peng, C., \& Im, M.\ 2007,
\apj, 658, 107
\bibitem[Letawe et al.(2009)]{lmcl} Letawe, G., Magain, P., Chantry, V., \& Letawe Y.\
2009, \mnras, 396, 78
\bibitem[Lauer(1999)]{l99} Lauer, T.~R.\ 1999, \pasp, 111, 227 
\bibitem[Lauer et al.(1995)]{l95}Lauer, T. R., Ajhar, E. A., Byun, Y.-I.,
Dressler, A., Faber, S. M., Grillmair,
C., Kormendy, J., Richstone, D., \& Tremaine, S. 1995, \aj, 110, 2622
\bibitem[Lucy(1974)]{lucy}Lucy, L. B. 1974, \aj, 79, 745
\bibitem[Magain et al.(2005)]{magain05}Magain, P., et al.\ 2005, \nat, 437, 381
\bibitem[Minkowski(1957)]{min} Minkowski, R.\ 1957, Radio astronomy, 4, 107
\bibitem[Richardson(1972)]{rich}Richardson, W. H. 1972, J. Opt. Soc. A., 62, 52
\bibitem[Sadler et al.(1989)]{sad} Sadler, E.~M., Jenkins, 
C.~R., \& Kotanyi, C.~G.\ 1989, \mnras, 240, 591 
\bibitem[Salviander et al.(2007)]{ssgb} Salviander, S., Shields, G. A., Gebhardt, K., 
\& Bonner, E. W.\ 2007, \apj, 662, 131
\bibitem[Schnittman \& Buonanno(2007)]{schnitt} Schnittman, J.~D., \& Buonanno, A.\ 2007, \apjl, 662, L63 
\bibitem[Shen et al.(2008)]{sgsrs} Shen, Y., Greene, J. E., Strauss, M. A., Richards,
G. T., \& Schneider, D. P. \ 2008, \apj, 680, 169
\bibitem[Strateva et al.(2003)]{strat} Strateva, I.~V., et 
al.\ 2003, \aj, 126, 1720 
\bibitem[Shuder(1982)]{shud} Shuder, J.~M.\ 1982, \apj, 259, 48 
\bibitem[Volonteri et al.(2003)]{vhm} Volonteri, M., 
Haardt, F., \& Madau, P.\ 2003, \apj, 582, 559 
\bibitem[Wiener(1949)]{wiener} Wiener, N. 1949, {\it Extrapolation,
Interpolation, and Smoothing of Stationary Time Series} (New York: Wiley)
\bibitem[Wrobel \& Heeschen(1991)]{wh} Wrobel, J.~M., \& Heeschen, D.~S.\ 1991,
\aj, 101, 148
\bibitem[Wrobel \& Laor(2009)]{wl} Wrobel, J.~M. \& Laor, A.\ 2009, \apjl,
in press, arXiv:0905.3566.
\end{thebibliography}
\end{document}